\documentclass[journal]{IEEEtran}
\ifCLASSINFOpdf

\else

\fi
\usepackage{amsmath}
\usepackage{makeidx}  
\usepackage{algorithm}
\usepackage{algorithmic}
\usepackage{graphicx}
\usepackage{subfigure}
\usepackage{epstopdf}
\usepackage{bm}
\usepackage{cite}
\usepackage{stfloats}

\usepackage{amssymb}
\usepackage{graphicx}

\usepackage{url}

\usepackage{tabularx,booktabs}
\newcolumntype{C}{>{\centering\arraybackslash}X} 
\usepackage{lipsum}

\usepackage{makecell} 

\usepackage{graphicx}
\usepackage{color}

\newtheorem{thm}{Theorem}

\newtheorem{pos}{Proposition}

\newtheorem{cor}{Corollary}

\begin{document}

\title{Static IRS Meets Distributed MIMO: A New Architecture for Dynamic Beamforming}

\author{Guangji~Chen,\IEEEmembership{}
        Qingqing~Wu,\IEEEmembership{}
        Celimuge~Wu,\IEEEmembership{}
        Mengnan~Jian,\IEEEmembership{}
        Yijian~Chen,\IEEEmembership{}
        and Wen~Chen\IEEEmembership{}\vspace{-20pt}

        \thanks{G. Chen is with the University of Macau, Macao 999078, China (email: guangjichen@um.edu.mo). Q. Wu and W. Chen are with Shanghai Jiao Tong University, 200240, China (e-mail: qingqingwu@sjtu.edu.cn; wenchen@sjtu.edu.cn). Celimuge Wu is with the University of Electro-Communications, Chofu 182-8585, Japan (e-mail: clmg@is.uec.ac.jp). M. Jian and Y. Chen are with ZTE Corporation, Shenzhen 518057, China (e-mail: jian.mengnan@zte.com.cn; chen.yijian@zte.com.cn).}}

\maketitle
\vspace{-24pt}
\begin{abstract}

Intelligent reflecting surface (IRS) has been considered as a revolutionary technology to enhance the wireless communication performance. To cater for multiple mobile users, adjusting IRS beamforming patterns over time, i.e., dynamic IRS beamforming (DIBF), is generally needed for achieving satisfactory performance, which results in high controlling power consumption and overhead. To avoid such cost, we propose a new architecture based on the static regulated IRS for wireless coverage enhancement, where the principle of distributed multiple-input multiple-output (D-MIMO) is integrated into the system to exploite the diversity of spatial directions provided by multiple access points (APs). For this new \emph{D-MIMO empowered static IRS} architecture, the total target area is partitioned into several subareas and each subarea is served by an assigned AP. We consider to maximize the worst-case received power over all locations in the target area by jointly optimizing a single set of IRS beamforming pattern and AP-subarea association. Then, a two-step algorithm is proposed to obtain its high-quality solution. Theoretical analysis unveils that the fundamental squared power gain can still be achieved over all locations in the target area. The performance gap relative to the DIBF scheme is also analytically quantified. Numerical results validate our theoretical findings and demonstrate the effectiveness of our proposed design over benchmark schemes.

\end{abstract}


\IEEEpeerreviewmaketitle

\vspace{-12pt}
\section{Introduction}
\vspace{-3pt}
Intelligent reflecting surface (IRS) has been envisioned as a promising technology to boost the spectrum efficiency and to enhance the coverage range for sixth-generation (6G) wireless networks \cite{di2020smart,8910627,chen2022irs}. Through altering the phase and/or amplitude of the impinging radio wave, IRSs are able to dynamically control the wireless channels for various objectives, such as signal enhancement, interference suppression, and blindness compensation \cite{wu2021intelligent}.
These promising benefits of IRSs have motivated intensive enthusiasm on integrating IRSs in various system setups and applications (see \cite{wu2021intelligent} and the references therein).

From the perspective of enhancing signal-to-noise ratio (SNR) of the wireless communication link, passive beamforming designs for users with fixed and known locations were investigated in \cite{wu2019beamforming, chen2023irs, zhi2021statistical, chen2023fundamental, mu2021capacity, wu2021irs, chen2022active}. In particular, the analytical result in the seminal work \cite{wu2019beamforming} demonstrated that the fundamental squared power gain of the IRS can be reaped even with the discrete phase shifts. In addition to providing high SNR, IRSs can also be deployed flexibly to create a virtual communication link between the access point (AP) and users in blind spots, thus ensuring the basic communication requirements of users located in the signal blind zone. The adoption of IRSs for coverage enhancement has been studied in previous works \cite{9462949, 9740570,9534744}, where the coverage range and system capacity were significantly improved.

However, all the above contributions rely on channel state information (CSI) of IRS involved links and dynamic configuration for IRS reflection phase-shifts, which incur significant signalling overhead. In the scenario where users change their positions frequently, the overhead of channel acquisition and dynamic configuring IRS phase-shifts may be unaffordable to cater for mobile users in real time. Moreover, exploiting the dynamic regulated IRS also leads to higher controlling power consumption compared to the static regulated IRS, which is more appealing for rapid deployment due to its advantage of simple control. Regarding the static regulated IRS, the works \cite{9351782, 9790792} investigated the static IRS beamforming design, where a single set of fixed IRS phase-shifts is designed to balance the passive beamforming gain in a target area. However, it has been demonstrated in \cite{wu2021irs} that employing static IRS beamforming suffers substantial performance-loss compared to the dynamic IRS beamforming (DIBF) scheme. Thus, one practical issue arises naturally: \emph{How to achieve high passive beamforming gain with a single set of fixed IRS reflection phase-shifts?}

\begin{figure}[!t]
\setlength{\abovecaptionskip}{-5pt}
\setlength{\belowcaptionskip}{-5pt}
\centering
\includegraphics[width= 0.3\textwidth]{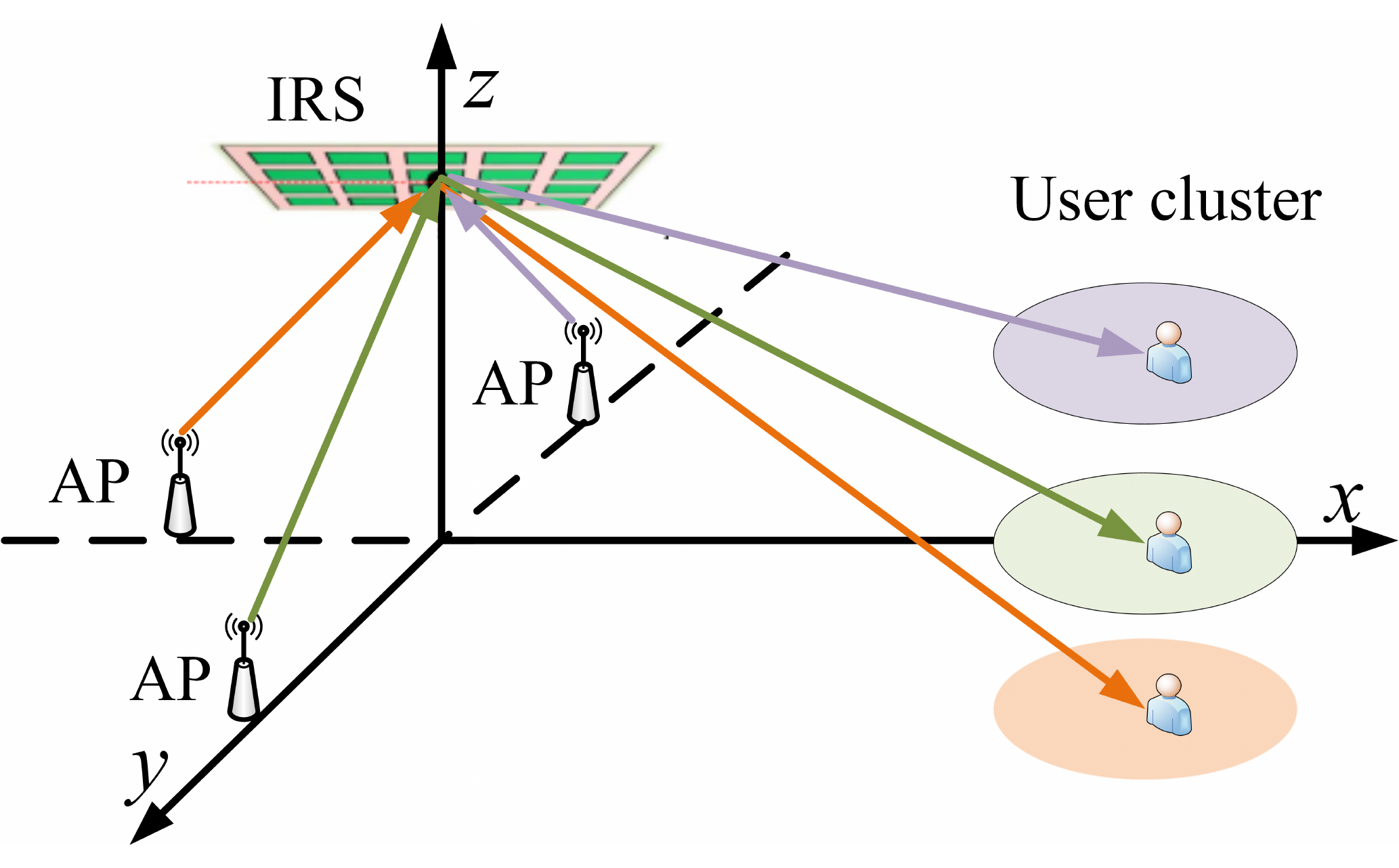}
\DeclareGraphicsExtensions.
\caption{D-MIMO empowered static IRS for coverage enhancement.}
\label{model}
\vspace{-20pt}
\end{figure}

To address the above issue, we propose a new architecture to unlock the DIBF gain for the static regulated IRS, where the concept of distributed multiple-input multiple-output (D-MIMO) is smartly integrated into the considered scenario. For the proposed architecture, namely \emph{D-MIMO empowered static IRS}, an IRS is deployed near a cluster of distributed APs to extend the signal coverage from APs to a given target area, as shown in Fig. 1. The target area is partitioned into several subareas and each subarea is served by an assigned AP in order to exploit the diversity of spatial directions provided by different APs. We aim to maximize the worst-case average received power over all locations in the target area by jointly
optimizing IRS beamforming and AP-subarea association. The formulated optimization problem is challenging to be solved since the IRS beamforming pattern is highly coupled with the AP-subarea association. Nevertheless, we propose an efficient two-step solution by decoupling the AP-subarea association optimization and IRS beamforming. Note that the proposed solution only relies on the location information and thus can be implemented off-line, which means that real-time CSI acquisition and dynamic control for IRS beamforming are not needed.
Then, by analytically characterizing the achievable objective value with respect to system parameters, we unveil that the fundamental squared power gain can still be achieved with only static IRS beamforming provided that the number of APs is sufficient. The performance gap relative to the DIBF scheme is also analytically quantified. Finally, we present simulation results to show the practical
advantages of the proposed architecture.
\vspace{-10pt}
\section{System model and Problem formulation}
As shown in Fig. 1, we introduce the \emph{D-MIMO empowered static IRS} architecture, where a cluster of distributed APs communicate with multiple single-antenna users with the aid of an $N$-element IRS. The number of the APs is denoted by $J$ and each AP is equipped with $M$ antennas. The users are uniformly distributed in a target area, denoted by ${\cal A}$. Due to the densely distributed obstacles, there are generally no line-of-sight (LoS) links between the APs and users and thus the direct links between the APs and users are assumed to be blocked. As a remedy, an IRS is deployed on the ceiling to establish virtual LoS links from the APs to the users. For convenience, we denote the sets of APs and reflecting elements of the IRS as ${\cal J} \buildrel \Delta \over = \left\{ {1, \ldots ,J} \right\}$ and ${\cal N} \buildrel \Delta \over = \left\{ {1, \ldots ,N} \right\}$, respectively. Note that in practice, it is difficult for the IRS to arbitrarily adjust its beamforming pattern to align served users due to the high controlling overhead/power consumption. To this end, we consider the static regulated IRS and thus use one set of fixed IRS beamforming pattern, denoted by ${\bf{\Theta }} = {\mathop{\rm diag}\nolimits} \left( {{e^{i{\theta _1}}}, \ldots ,{e^{i{\theta _N}}}} \right)$, where ${\theta _n} \in \left( {0,2\pi } \right]$ denotes the phase-shift of the $n$-th element of the IRS, $n \in {\cal N}$. To improve the received signal power in the target area, we propose the AP selection protocol for the \emph{D-MIMO empowered static IRS} architecture, described as follows.

The target area ${\cal A}$ is further partitioned into $K$ disjoint sub-areas, denoted by ${{\cal A}_k}$, $k \in {\cal K} \buildrel \Delta \over = \left\{ {1, \ldots ,K} \right\}$, which satisfy $\bigcup\nolimits_{k \in {\cal K}} {{{\cal A}_k}}  = {\cal A}$ and ${{\cal A}_p}\bigcap {{{\cal A}_q}}  = \emptyset$, $\forall p \ne q,p,q \in {\cal K}$. Each sub-area ${{{\cal A}_k}}$ can be associated with one AP $j$, $j \in {\cal J}$ and users in sub-area ${{{\cal A}_k}}$ will receive its desired signal from AP $j$. Thus, we introduce the set of binary variables $\left\{ {{\lambda _{k,j}}} \right\}$, $k \in {\cal K}$, $j \in {\cal J}$, which indicates that sub-area ${{{\cal A}_k}}$ is associated with AP $j$ if ${\lambda _{k,j}} = 1$; otherwise, ${\lambda _{k,j}} = 0$. To simplify the practical implementation, we assume that the user in each sub-area can only be associated with at most one AP in ${\cal J}$. Thus, we have $\sum\nolimits_{j = 1}^J {{\lambda _{k,j}}}  \le 1$. Specifically, the users in the target area ${\cal A}$ are scheduled following the round robin scheme and the scheduled user in a specific time slot is referred to as the typical user. When the typical user is in sub-area ${{{\cal A}_k}}$, AP $j$ in ${\cal J}$ which satisfies ${\lambda _{k,j}} = 1$ would be activated to serve the typical user. The remaining APs in ${\cal J}/\left\{ j \right\}$ are available to serve users outside the target area by adjusting their beam directions, which may not generate strong interference to the typical user. Note that the controlling overhead for determining $\left\{ {{\lambda _{k,j}}} \right\}$ scales linearly with respect to $J$, whereas the associated overhead for implementing the DIBF scheme is proportional to $N$. Since $J \ll N$, the controlling overhead is significantly reduced.

The base-band channels from AP $j$ to the IRS and from the IRS to the typical user in sub-area ${{{\cal A}_k}}$ are denoted by ${{\bf{G}}_j} \in \mathbb{C}^{N\times M}$ and ${{\bf{h}}_k}\left( {{{\bf{u}}_{{k}}}} \right) \in \mathbb{C}^{N\times 1}$, where ${{\bf{u}}_{{k}}} \in {{\cal A}_k}$ denotes the position of the typical user in sub-area ${{\cal A}_k}$. Without loss of generality, we employ the Rician fading model to characterize the AP $j$-IRS and IRS-${{{\bf{u}}_{{k}}}}$ links as
\begin{align}\label{channel}
&{{\bf{h}}_k}\left( {{{\bf{u}}_{{k}}}} \right) \!\!=\!\! {\rho _{2,k}}\left( {{{\bf{u}}_{{k}}}} \right)\left( {\sqrt {\frac{\varepsilon }{{\varepsilon  \!+\! \varepsilon }}} {{{\bf{\bar h}}}_k}\left( {{{\bf{u}}_{{k}}}} \right) \!\!+\!\! \sqrt {\frac{1}{{\varepsilon  \!+\! 1}}} {{{\bf{\tilde h}}}_k}\left( {{{\bf{u}}_{{k}}}} \right)} \right)\!,\\
&{{\bf{G}}_j} = {\rho _{1,j}}\left( {\sqrt {\frac{\delta }{{\delta  + 1}}} {{{\bf{\bar G}}}_j} + \sqrt {\frac{1}{{\delta  + 1}}} {{{\bf{\tilde G}}}_j}} \right),
\end{align}
where $\rho _{2,k}^2\left( {{{\bf{u}}_{{k}}}} \right)$ and $\rho _{1,j}^2$ are distance-based large-scale path-loss, $\varepsilon $ and $\delta $ are Rician factors. ${{{{\bf{\bar h}}}_k}\left( {{{\bf{u}}_{{k}}}} \right)}$ and ${{{{\bf{\bar G}}}_j}}$ are deterministic LoS channel components. By contrast, ${{{{\bf{\tilde h}}}_k}\left( {{{\bf{u}}_{{k}}}} \right)}$ and ${{{{\bf{\tilde G}}}_j}}$ are non-LoS (NLoS) channel components, whose elements are independent and identical distribution random variables following ${\cal C}{\cal N}\left( {0,1} \right)$. For ease of exposition, we assume that both the APs and the IRS adopt uniform linear arrays. Then, ${{{{\bf{\bar G}}}_j}}$ and ${{{{\bf{\bar h}}}_k}\left( {{{\bf{u}}_{{k}}}} \right)}$ can be respectively expressed as ${{{\bf{\bar G}}}_j} = {{\bf{a}}_N}\left( {\phi _{r,j}^z,\eta _{r,j}^a} \right){\bf{a}}_M^H\left( {\phi _{t,j}^z,\eta _{t,j}^a} \right)$, $j \in {\cal J}$, and ${{{\bf{\bar h}}}_k}\left( {{{\bf{u}}_{{k}}}} \right) = {{\bf{a}}_N}\left( {\phi _{t,k}^z\left( {{{\bf{u}}_{{k}}}} \right),\eta _{t,k}^a\left( {{{\bf{u}}_{{k}}}} \right)} \right)$, ${{\bf{u}}_{{k}}} \in {{\cal A}_k}$, $k \in {\cal K}$, with ${{\bf{a}}_N}\left( {\phi ,\eta } \right) = \left[ {1, \ldots ,{e^{ - i2\pi \left( {N - 1} \right)\frac{d}{\lambda }\sin \phi \cos \eta }}} \right]$,
where $d$ is the elements/antennas spacing, $\lambda$ represents wavelength, ${\phi _{t,j}^z}$, ${\eta _{t,j}^a}$ (${\phi _{t,k}^z\left( {{{\bf{u}}_{{k}}}} \right)}$, ${\eta _{t,k}^a\left( {{{\bf{u}}_{{k}}}} \right)}$) are respectively the zenith and azimuth angles of departure (AoD) from AP $j$ to the IRS (from the IRS to the user located at ${{{\bf{u}}_{{k}}}}$). ${\phi _{r,j}^z}$, ${\eta _{r,j}^a}$ are respectively the zenith and azimuth angles of arrival (AoA) from AP $j$ to the IRS.

The transmit beamforming vector of AP $j$ is denoted by ${{\bf{w}}_j}$. Then, the received signal power of the typical user located at ${{{\bf{u}}_{{k}}}}$ can be expressed as
\begin{align}\label{received_power}
{p_r}\left( {{{\bf{u}}_{{k}}},{{\cal A}_k},{\lambda _{k,j}}} \right) = \sum\nolimits_{j = 1}^J {{\lambda _{k,j}}} {\left| {{\bf{h}}_k^H\left( {{{\bf{u}}_{{k}}}} \right){\bf{\Theta }}{{\bf{G}}_j}{{\bf{w}}_j}} \right|^2}.
\end{align}
Our objective is to maximize the worst-case/minimum average received power within the target area ${\cal A}$, by jointly optimizing the static IRS beamforming pattern ${\bf{\Theta }}$, subarea-AP association $\left\{ {{\lambda _{k,j}}} \right\}$, and the transmit beamforming at the AP ${{\bf{w}}_j}$. The maximum transmit power of APs is denoted by ${P_{\max }}$ and the optimization problem is formulated as
\begin{subequations}\label{C1}
\begin{align}
\label{C1-a}\mathop {\max }\limits_{{\bf{\Theta }},\left\{ {{\lambda _{k,j}}} \right\}} & \;\;\mathop {\min }\limits_{k \in {\cal K},{{\bf{u}}_k} \in {{\cal A}_k}} {\mathop{\rm E}\nolimits} \left[ {\mathop {\max }\limits_{\left\{ {{{\bf{w}}_j}} \right\}}\;\; {p_r}\left( {{{\bf{u}}_k},{{\cal A}_k},{\lambda _{k,j}}} \right)} \right]\\
\label{C1-b}{\rm{s.t.}}\;\;\;&\sum\nolimits_{j = 1}^J {{\lambda _{k,j}}}  \le 1,\forall k,\\
\label{C1-c}&{{\lambda _{k,j}}} \in \left\{ {0,1} \right\},\forall k,j,\\
\label{C1-d}&{\left\| {{{\bf{w}}_j}} \right\|^2} \le {P_{\max }},\forall j,\\
\label{C1-e}&\left| {{{\left[ {\bf{\Theta }} \right]}_{n,n}}} \right| = 1,\forall n.
\end{align}
\end{subequations}
Notice that solving problem \eqref{C1} is very challenging in general. First, the objective function is difficult to be expressed in terms of optimization variables since it is the worst-case received power over a two-dimensional area. Second, the optimization variables $\left\{ {{{\bf{w}}_j}} \right\}$, $\left\{ {{\lambda _{k,j}}} \right\}$ and ${\bf{\Theta }}$ are intricately coupled in the objective function.
\vspace{-10pt}
\section{Proposed Solution and Performance Analysis}
In this section, we first propose an efficient algorithm to obtain high-quality sub-optimal solution of problem \eqref{C1}. To gain more engineering insights, we further theoretically characterize the worst-case received power in terms of system parameters, which is useful for quantifying the performance-loss between our proposed architecture and the DIBF scheme.
\vspace{-10pt}
\subsection{Proposed Solution to Problem \eqref{C1}}
To solve problem \eqref{C1}, we first consider the optimization of $\left\{ {{{\bf{w}}_j}} \right\}$. For any given $\left\{ {{\lambda _{k,j}}} \right\}$ and ${\bf{\Theta }}$, it can be readily shown that the optimal transmit beamforming vector ${{{\bf{w}}_j}}$ follows the maximum ratio transmission towards the equivalent channel ${{\bf{h}}_k^H\left( {{{\bf{u}}_{{k}}}} \right){\bf{\Theta }}{{\bf{G}}_j}}$ of the typical user, i.e., ${\bf{w}}_j^* = {P_{\max }}{\left( {{\bf{h}}_k^H\left( {{{\bf{u}}_{{k}}}} \right){\bf{\Theta }}{{\bf{G}}_j}} \right)^H}/\left\| {{\bf{h}}_k^H\left( {{{\bf{u}}_{{k}}}} \right){\bf{\Theta }}{{\bf{G}}_j}} \right\|$. To this end, we have the following proposition.
\begin{pos}
Problem \eqref{C1} is equivalent to
\begin{subequations}\label{C2}
\begin{align}
\label{C2-a}\mathop {\max }\limits_{{\bf{\Theta }}, \left\{ {{\lambda _{k,j}}} \right\}} & \;\;\mathop {\min }\limits_{k \in {\cal K}} \mathop {\min }\limits_{{{\bf{u}}_{{k}}} \in {{\cal A}_k}} \sum\nolimits_{j = 1}^J {{\lambda _{k,j}}} \bar p_r^{k,j}\left( {{{\bf{u}}_{{k}}},{A_k},{\lambda _{k,j}}} \right)\\
\label{C2-a}{\rm{s.t.}}\;\;\;\;&\eqref{C1-b}, \eqref{C1-c}, \eqref{C1-e},
\end{align}
\end{subequations}
where
\begin{align}\label{average_received_power}
&\bar p_r^{k,j}\left( {{{\bf{u}}_{{k}}},{A_k},{\lambda _{k,j}}} \right)\nonumber\\
&= {P_{\max }}\rho _{1,j}^2\rho _{2,k}^2\left( {{{\bf{u}}_{{k}}}} \right)\left( {{\gamma _1}{{\left\| {{\bf{\bar h}}_k^H\left( {{{\bf{u}}_k}} \right){\bf{\Theta }}{{{\bf{\bar G}}}_j}} \right\|}^2} \!\!+\!\! {\gamma _2}MN} \right),
\end{align}
with ${\gamma _1} = \varepsilon \delta /\left( {\left( {\varepsilon  + 1} \right)\left( {\delta  + 1} \right)} \right)$ and ${\gamma _2} = 1 - {\gamma _1}$.
\end{pos}

\emph{Proof:} Plugging the optimal transmit beamforming vector ${\bf{w}}_j^* = {P_{\max }}{\left( {{\bf{h}}_k^H\left( {{{\bf{u}}_{{k}}}} \right){\bf{\Theta }}{{\bf{G}}_j}} \right)^H}/\left\| {{\bf{h}}_k^H\left( {{{\bf{u}}_{{k}}}} \right){\bf{\Theta }}{{\bf{G}}_j}} \right\|$ into the objective function \eqref{C2-a}, we have
\begin{align}\label{average_received_power1}
&{\rm{E}}\left[ {\mathop {\max }\limits_{\left\{ {{{\bf{w}}_j}} \right\}} \mathop {\min }\limits_{k \in {\cal K}} \mathop {\min }\limits_{{{\bf{u}}_{{k}}} \in {{\cal A}_k}} {p_r}\left( {{{\bf{u}}_{{k}}},{A_k},{\lambda _{k,j}}} \right)} \right]\nonumber\\
& = \mathop {\min }\limits_{k \in {\cal K}} \mathop {\min }\limits_{{{\bf{u}}_{{k}}} \in {{\cal A}_k}} {P_{\max }}\sum\nolimits_{j = 1}^J {{\lambda _{k,j}}} {\rm{E}}\left[ {{{\left\| {{\bf{h}}_k^H\left( {{{\bf{u}}_k}} \right){\bf{\Theta }}{{\bf{G}}_j}} \right\|}^2}} \right].
\end{align}
Then, we focus on deriving ${\rm{E}}\left[ {{{\left\| {{\bf{h}}_k^H\left( {{{\bf{u}}_k}} \right){\bf{\Theta }}{{\bf{G}}_j}} \right\|}^2}} \right]$ as follows
\begin{align}\label{expectation}
&{\rm{E}}\left[ {{{\left\| {{\bf{h}}_k^H\left( {{{\bf{u}}_k}} \right){\bf{\Theta }}{{\bf{G}}_j}} \right\|}^2}} \right]\nonumber\\
& = \rho _{1,j}^2\rho _{2,k}^2\left( {{{\bf{u}}_{{k}}}} \right)\left( {{\gamma _1}{{\left\| {{\bf{\bar h}}_k^H\left( {{{\bf{u}}_k}} \right){\bf{\Theta }}{{{\bf{\bar G}}}_j}} \right\|}^2} + \sum\nolimits_{m = 1}^3 {{\mathop{\rm E}\nolimits} \left[ {{{\left\| {{{\bf{z}}_m}} \right\|}^2}} \right]} } \right)\nonumber\\
& \mathop  = \limits^{\left( a \right)}  \rho _{1,j}^2\rho _{2,k}^2\left( {{{\bf{u}}_{{k}}}} \right)\left( {{\gamma _1}{{\left\| {{\bf{\bar h}}_k^H\left( {{{\bf{u}}_k}} \right){\bf{\Theta }}{{{\bf{\bar G}}}_j}} \right\|}^2} + {\gamma _2}MN} \right),
\end{align}
where
\begin{align}\label{notation1}
&{{\bf{z}}_1} = \delta {\bf{\tilde h}}_k^H\left( {{{\bf{u}}_k}} \right){\bf{\Theta }}{{{\bf{\bar G}}}_j}/\left( {\left( {\varepsilon  + 1} \right)\left( {\delta  + 1} \right)} \right),\nonumber\\
&{{\bf{z}}_2} = \varepsilon {\bf{\bar h}}_k^H\left( {{{\bf{u}}_k}} \right){\bf{\Theta }}{{{\bf{\tilde G}}}_j}/\left( {\left( {\varepsilon  + 1} \right)\left( {\delta  + 1} \right)} \right),\nonumber\\
&{{\bf{z}}_3} = {\bf{\tilde h}}_k^H\left( {{{\bf{u}}_k}} \right){\bf{\Theta \tilde G}}/\left( {\left( {\varepsilon  + 1} \right)\left( {\delta  + 1} \right)} \right),
\end{align}
and (a) follows from
\begin{align}\label{expectation1}
&{\mathop{\rm E}\nolimits} \left[ {{{\left\| {{\bf{\tilde h}}_k^H\left( {{{\bf{u}}_k}} \right){\bf{\Theta \bar G}}} \right\|}^2}} \right] = {\mathop{\rm E}\nolimits} \left[ {{{\left\| {{\bf{\bar h}}_k^H\left( {{{\bf{u}}_k}} \right){\bf{\Theta }}{{{\bf{\tilde G}}}_j}} \right\|}^2}} \right] \nonumber\\
&= {\mathop{\rm E}\nolimits} \left[ {{{\left\| {{\bf{\tilde h}}_k^H\left( {{{\bf{u}}_k}} \right){\bf{\Theta \tilde G}}} \right\|}^2}} \right] = MN.
\end{align}
By applying \eqref{expectation} into \eqref{average_received_power1}, the proof is completed.
$\hfill\blacksquare$

Proposition 1 indicates that the objective function \eqref{C2-a} is independent of the fast-varying instantaneous CSI ${{{{\bf{\tilde G}}}_j}}$ and ${{{\bf{\tilde h}}}_k}\left( {{{\bf{u}}_k}} \right)$. Instead, it only relies on the statistical CSI, i.e., AoD and AoA in ${{{{\bf{\bar G}}}_j}}$, the path-loss coefficients $\rho _{1,j}^2$, $\rho _{2,k}^2\left( {{{\bf{u}}_{{k}}}} \right)$, and Rician factors $\delta$, $\varepsilon $. Note that the corresponding statistical CSI only depends on the locations of the distributed APs, IRS, and the target area, which remains static over the time. As such, the optimization of ${\bf{\Theta }}$ and $\left\{ {{\lambda _{k,j}}} \right\}$ can be implemented offline and the optimized results are not required to be varied according to instantaneous CSI, which effectively eases the synchronization, control, and channel estimation requirements in practical systems. Problem \eqref{C2} is still challenging to be solved due to the coupled optimization variables $\left\{ {{\lambda _{k,j}}} \right\}$ and ${\bf{\Theta }}$ in the objective function. By fully exploiting the special structure of problem \eqref{C2}, $\bar p_r^{k,j}\left( {{{\bf{u}}_{{k}}},{A_k},{\lambda _{k,j}}} \right)$ in its objective function \eqref{C2-a} can be re-expressed as follows
\begin{align}\label{average_received_power}
&\bar p_r^{k,j}\left( {{{\bf{u}}_{{k}}},{A_k},{\lambda _{k,j}}} \right)\nonumber\\
& = {P_{\max }}\rho _{1,j}^2\rho _{2,k}^2\left( {{{\bf{u}}_{{k}}}} \right)\left( {{\gamma _1}{{\left| {{\chi _{k,j}}} \right|}^2}{{\left\| {{\bf{a}}_M^H\left( {\phi _{t,j}^z,\eta _{t,j}^a} \right)} \right\|}^2} + {\gamma _2}MN} \right)\nonumber\\
& = {P_{\max }}\rho _{1,j}^2\rho _{2,k}^2\left( {{{\bf{u}}_{{k}}}} \right)\left( {{\gamma _1}{{\left| {{\chi _{k,j}}} \right|}^2}M + {\gamma _2}MN} \right),
\end{align}
where ${\chi _{k,j}}{\rm{ = }}{\bf{a}}_N^H\left( {\phi _{t,k}^z\left( {{{\bf{u}}_{{k}}}} \right),\eta _{t,k}^a\left( {{{\bf{u}}_{{k}}}} \right)} \right){\bf{\Theta }}{{\bf{a}}_N}\left( {\phi _{r,j}^z,\eta _{r,j}^a} \right)$. We define ${{{\left| {{\chi _{k,j}}} \right|}^2}}$ as the passive beamforming gain of the IRS when ${{\bf{u}}_{{k}}} \in {{\cal A}_k}$ is associated with AP $j$, i.e., ${\lambda _{k,j}} = 1$ . ${\left| {{\chi _{k,j}}} \right|^2}$ can be further expressed as
\begin{align}\label{passive_BF_gain}
{\left| {{\chi _{k,j}}} \right|^2} = {\left| {\sum\nolimits_{n = 1}^N {{e^{j\left( {{\theta _n} + 2\pi \left( {n - 1} \right)\bar d\left( {{\Phi _{t,k}}\left( {{{\bf{u}}_{{k}}}} \right) - {\Omega _{r,j}}} \right)} \right)}}} } \right|^2},
\end{align}
where ${\Phi _{t,k}}\left( {{{\bf{u}}_{{k}}}} \right) = \sin \phi _{t,k}^z\left( {{{\bf{u}}_{{k}}}} \right)\cos \eta _{t,k}^a\left( {{{\bf{u}}_{{k}}}} \right)$, ${\Omega _{r,j}} = \sin \phi _{r,j}^z\cos \eta _{r,j}^a$, and $\bar d = d/\lambda$. By ignoring the constant term,  i.e., ${{\gamma _2}MN}$, in \eqref{average_received_power}, we transform problem \eqref{C2} in an equivalent form as
\begin{subequations}\label{C3}
\begin{align}
\label{C3-a}\mathop {\max }\limits_{{\bf{\Theta }}, \left\{ {{\lambda _{k,j}}} \right\}} & \;\;\mathop {\min }\limits_{k \in {\cal K}} \mathop {\min }\limits_{{{\bf{u}}_{{k}}} \in {{\cal A}_k}} \sum\nolimits_{j = 1}^J {{\lambda _{k,j}}} \rho _{1,j}^2\rho _{2,k}^2\left( {{{\bf{u}}_{{k}}}} \right){\left| {{\chi _{k,j}}} \right|^2}\\
\label{C3-a}{\rm{s.t.}}\;\;\;\;&\eqref{C1-b}, \eqref{C1-c}, \eqref{C1-e}.
\end{align}
\end{subequations}

In the following, we focus on solving problem \eqref{C3}. To avoid that the achieved passive beamforming gain is overwhelmed by distinct path-loss of different APs-IRS links, we assume that $\rho _{1,1}^2 =  \ldots  = \rho _{1,J}^2 = \rho _1^2$, which can be easily realized by deploying the APs at the positions with the similar distance to the IRS. By ignoring $\rho _{2,k}^2\left( {{{\bf{u}}_{{k}}}} \right)$ in the inner minimization of \eqref{C3-a}, we transform problem \eqref{C3} into an approximate one as
\begin{subequations}\label{C4}
\begin{align}
\label{C4-a}\mathop {\max }\limits_{{\bf{\Theta }}, \left\{ {{\lambda _{k,j}}} \right\}} & \;\;\mathop {\min }\limits_{k \in {\cal K}} \mathop {\min }\limits_{{{\bf{u}}_{{k}}} \in {{\cal A}_k}} \sum\nolimits_{j = 1}^J {{\lambda _{k,j}}} {\left| {{\chi _{k,j}}} \right|^2}\\
\label{C4-a}{\rm{s.t.}}\;\;\;\;&\eqref{C1-b}, \eqref{C1-c}, \eqref{C1-e}.
\end{align}
\end{subequations}
This approximation is reasonable since the passive beamforming gain ${\left| {{\chi _{k,j}}} \right|^2}$ is more sensitive than $\rho _{2,k}^2\left( {{{\bf{u}}_{{k}}}} \right)$ to the location variation of ${{{\bf{u}}_{{k}}}}$ in subarea ${{\cal A}_k}$. Based on the special structure of problem \eqref{C4}, we propose an efficient sub-optimal solution by decoupling ${\bf{\Theta }}$ and $\left\{ {{\lambda _{k,j}}} \right\}$. For arbitrarily given $\left\{ {{\lambda _{k,j}}} \right\}$, the optimization of ${\bf{\Theta }}$ is equivalent to maximize the worst-case passive beamforming gain within a given angular span, which can be written as
\begin{subequations}\label{C5}
\begin{align}
\label{C5-a}\mathop {\max }\limits_{\bf{\Theta }}& \mathop {\min }\limits_{{\Delta _{\min }}\left( {\left\{ {{\lambda _{k,j}}} \right\}} \right) \! \le \! \Delta  \! \le \! {\Delta _{\max }}\left( {\left\{ {{\lambda _{k,j}}} \right\}} \right)} {\left| {\sum\limits_{n = 1}^N {{e^{i\left( {{\theta _n} \!+\! 2\pi \left( {n \!-\! 1} \right)\bar d\Delta } \right)}}} } \right|^2}\\
\label{C5-a}{\rm{s.t.}}\;&0 \le {\theta _n} < 2\pi, \forall n \in {\cal N},
\end{align}
\end{subequations}
where
\begin{align}\label{max_min_angular}
&{\Delta _{\max }}\left( {\left\{ {{\lambda _{k,j}}} \right\}} \right) \!=\! \mathop {\max }\limits_{k \in {\cal K}} \sum\limits_{j = 1}^J {{\lambda _{k,j}}\mathop {\max }\limits_{{{\bf{u}}_{{k}}} \in {{\cal A}_k}} \left( {{\Phi _{t,k}}\left( {{{\bf{u}}_{{k}}}} \right) - {\Omega _{r,j}}} \right)},\\
&{\Delta _{\min }}\left( {\left\{ {{\lambda _{k,j}}} \right\}} \right) = \mathop {\min }\limits_{k \in {\cal K}} \sum\limits_{j = 1}^J {{\lambda _{k,j}}\mathop {\min }\limits_{{{\bf{u}}_{{k}}} \in {{\cal A}_k}} \left( {{\Phi _{t,k}}\left( {{{\bf{u}}_{{k}}}} \right) - {\Omega _{r,j}}} \right)}.
\end{align}
We define ${\Delta _s}\left( {\left\{ {{\lambda _{k,j}}} \right\}} \right) \buildrel \Delta \over = {\Delta _{\max }}\left( {\left\{ {{\lambda _{k,j}}} \right\}} \right) - {\Delta _{\min }}\left( {\left\{ {{\lambda _{k,j}}} \right\}} \right)$ as the angular deviation associated with ${\left\{ {{\lambda _{k,j}}} \right\}}$. Then, we have the following proposition.
\begin{pos}
The optimal objective value of problem \eqref{C5} is non-increasing with respect to ${\Delta _s}\left( {\left\{ {{\lambda _{k,j}}} \right\}} \right)$.
\end{pos}

\emph{Proof:} Without loss of generality, two angular spans $\left[ {{\Delta _a},{\Delta _b}} \right]$ and $\left[ {{\Delta _a},{\Delta _c}} \right]$ with ${\Delta _c} \ge {\Delta _b}$ are considered. As such, their associated angular deviations are $\Delta _s^1 = {\Delta _b} - {\Delta _a}$ and $\Delta _s^2 = {\Delta _c} - {\Delta _a}$, respectively. $\Delta _s^2 \ge \Delta _s^1$ and $\left[ {{\Delta _a},{\Delta _c}} \right] = \left[ {{\Delta _a},{\Delta _b}} \right] \cup \left[ {{\Delta _b},{\Delta _c}} \right]$. Suppose that the optimal IRS phase-shift to maximize $\mathop {\min }\limits_{{\Delta _a} \le \Delta  \le {\Delta _c}} {\left| {\sum\nolimits_{n = 1}^N {{e^{i\left( {{\theta _n} + 2\pi \left( {n - 1} \right)\bar d\Delta } \right)}}} } \right|^2}$ is ${{\bf{\Theta }}^*} = {\rm{diag}}\left( {{e^{i\theta _1^*}}, \ldots ,{e^{i\theta _N^*}}} \right)$ and the corresponding objective value is ${g_2}\left( {\Delta _s^2} \right)$. Then, we have
\begin{align}\label{inequality1}
{g_2}\left( {\Delta _s^2} \right) &\!=\! \min \left\{ \begin{array}{l}
\mathop {\min }\limits_{{\Delta _a} \le \Delta  \le {\Delta _b}} {\left| {\sum\nolimits_{n = 1}^N {{e^{i\left( {\theta _n^* + 2\pi \left( {n - 1} \right)\bar d\Delta } \right)}}} } \right|^2},\\
\mathop {\min }\limits_{{\Delta _b} \le \Delta  \le {\Delta _c}} {\left| {\sum\nolimits_{n = 1}^N {{e^{i\left( {\theta _n^* + 2\pi \left( {n - 1} \right)\bar d\Delta } \right)}}} } \right|^2}
\end{array} \right\},\nonumber\\
& \!\le\! \mathop {\min }\limits_{{\Delta _a} \le \Delta  \le {\Delta _b}} {\left| {\sum\limits_{n = 1}^N {{e^{i\left( {\theta _n^* + 2\pi \left( {n - 1} \right)\bar d\Delta } \right)}}} } \right|^2}\!\!\mathop  \le \limits^{\left( a \right)} \!\!{g_1}\left( {\Delta _s^2} \right),
\end{align}
where (a) follows by that ${{\bf{\Theta }}^*}$ is one feasible solution to maximize $\mathop {\min }\limits_{{\Delta _a} \le \Delta  \le {\Delta _b}} {\left| {\sum\nolimits_{n = 1}^N {{e^{i\left( {{\theta _n} + 2\pi \left( {n - 1} \right)\bar d\Delta } \right)}}} } \right|^2}$ and ${g_1}\left( {\Delta _s^2} \right)$ is its optimal objective value. Thus, we completes the proof.
$\hfill\blacksquare$

Proposition 2 identifies the fact that the optimization of $\left\{ {{\lambda _{k,j}}} \right\}$ in \eqref{C4} is equivalently to minimize the resulting angular deviation, which leads to the optimization problem as
\begin{align}\label{association}
\mathop {\min }\limits_{\left\{ {{\lambda _{k,j}}} \right\}} {\Delta _s}\left( {\left\{ {{\lambda _{k,j}}} \right\}} \right)\;\;\;\; {\rm{s.t.}}\;\;\eqref{C1-b}, \eqref{C1-c}.
\end{align}
Note that the angular deviation can be effectively adjusted by switching the AP association to control ${\Delta _{\max }}\left( {\left\{ {{\lambda _{k,j}}} \right\}} \right)$ and ${\Delta _{\min }}\left( {\left\{ {{\lambda _{k,j}}} \right\}} \right)$, which provides us an efficient successive refinement algorithm to reduce the angular deviation iteratively. The details are shown as in Algorithm 1.
\begin{algorithm}[!t]\label{method1}
\caption{Successive Refinement Algorithm for solving \eqref{association}}
\begin{algorithmic}[1]
\STATE Initialize ${\left\{ {{\lambda _{k,j}}} \right\}}$ by assigning all the sub-areas to AP $j$ with $\arg \mathop {\min }\limits_{j \in {\cal J}} {\Omega _{r,j}}$.
\STATE {\bf repeat}
\STATE Determine the bottleneck sub-area ${{\cal A}_{{k_b}}}$ as ${\mathop {\max }\limits_{{{\bf{u}}_{{k}}} \in {{\cal A}_k}} \left( {{\Phi _{t,k}}\left( {{{\bf{u}}_{{k}}}} \right) - {\Omega _{r,j}}} \right)}$ and assign AP ${j^*}$ to subarea ${{\cal A}_{{k_b}}}$ which results in the minimum value of ${\Delta _s}\left( {\left\{ {{\lambda _{k,j}}} \right\}} \right)$.
\STATE {\bf until} the objective value ${\Delta _s}\left( {\left\{ {{\lambda _{k,j}}} \right\}} \right)$ can not be decreased.
\end{algorithmic}
\end{algorithm}

With the obtained ${{\lambda _{k,j}}}$ by Algorithm 1, the IRS phase-shift ${\bf{\Theta }}$ is optimized by solving problem \eqref{C5}. Such a problem can be solved by adopting the beam flattening technique proposed in \cite{9351782} and the details are omitted for brevity.
\vspace{-8pt}
\subsection{Theoretical Analysis}
In this subsection, we provide theoretical analysis to characterize the worst-case passive beamforming gain, which is useful to quantify the practical performance gap between the proposed architecture and the DIBF scheme. First, we aim to measure the ability of using distributed APs to reduce the corresponding angular deviation. Let $\Delta _s^I \buildrel \Delta \over = \Delta _{\max }^I - \Delta _{\min }^I$ denote the initial angular deviation by assigning one AP (e.g., AP 1) to all the subareas, where $\Delta _{\max }^I = {\Delta _{\max }}\left\{ {\lambda _{k,j}^I} \right\}$ and $\Delta _{\min }^I = {\Delta _{\min }}\left\{ {\lambda _{k,j}^I} \right\}$ with $\lambda _{k,1}^I = 1,\forall k,\lambda _{k,j}^I = 0,j \ne 1,\forall k$. Note that the initial angular deviation depends on the locations of the total area ${\cal A}$. Then, we have the following proposition.
\begin{pos}
Under the condition that ${\Phi _{r,j}} = {\Phi _{r,1}} + \left( {j - 1} \right)\Delta _s^I/J$, the optimal objective value of problem \eqref{max_min_angular} is $\Delta _s^* = \Delta _s^I/J$.
\end{pos}

\emph{Proof:} We prove $\Delta _s^* = \Delta _s^I/J$ by showing that $\Delta _s^* \ge \Delta _s^I/J$ and $\Delta _s^* \le \Delta _s^I/J$, respectively. First, we focus on proving $\Delta _s^* \ge \Delta _s^I/J$ as follows. The lower bound of the value of ${\Delta _{\max }}\left\{ {{\lambda _{k,j}}} \right\}$ can be obtained as $\Delta _{\max }^{lb}\left\{ {{\lambda _{k,j}}} \right\} = \Delta _{\max }^I + {\Phi _{r,1}} - {\Phi _{r,J}}$. Similarly, the upper bound of the value of ${\Delta _{\min }}\left\{ {{\lambda _{k,j}}} \right\}$ is $\Delta _{\min }^{ub}\left\{ {{\lambda _{k,j}}} \right\} = \Delta _{\min }^I$. As such, we have $\Delta _s^* \ge \Delta _{\max }^{lb}\left\{ {{\lambda _{k,j}}} \right\} - \Delta _{\min }^{ub}\left\{ {{\lambda _{k,j}}} \right\} = \Delta _s^I - \left( {j - 1} \right)\Delta _s^I/J = \Delta _s^I/J$. Then, we focus on proving $\Delta _s^* \le \Delta _s^I/J$ by demonstrating that $\Delta _s^I/J$ is achievable. Specifically, ${\cal A}$ can be partitioned into $K = J$ subareas. Let $\Phi _t^{\min } = \Delta _{\min }^I + {\Phi _{r,1}}$ and $\Phi _t^{\max } = \Delta _{\max }^I + {\Phi _{r,1}}$. For each subarea ${{\cal A}_j}$, it satisfies ${\Phi _{t,j}}\left( {{{\bf{u}}_j}} \right) \in \left[ {\Phi _t^{\min } + \left( {j - 1} \right)\Delta _s^I/J,\Phi _t^{\min } + j\Delta _s^I/J} \right]$, ${{\bf{u}}_j} \in {{\cal A}_j}$. By assigning AP $j$ to subarea ${{\cal A}_j}$, i.e., ${\lambda _{j,j}} = 1$. We have ${\Delta _{\max }}\left\{ {{\lambda _{k,j}}} \right\} - {\Delta _{\min }}\left\{ {{\lambda _{k,j}}} \right\} = \Delta _s^I/J$ and thus $\Delta _s^* \le \Delta _s^I/J$. Based on the facts that $\Delta _s^* \ge \Delta _s^I/J$ and $\Delta _s^* \le \Delta _s^I/J$, we obtain $\Delta _s^* = \Delta _s^I/J$, which thus completes the proof.
$\hfill\blacksquare$

Proposition 3 explicitly unveils that the corresponding angular deviation can be effectively reduced from $\Delta _s^I$ to $\Delta _s^I/J$ by properly designing the AP-subarea associations. Define ${\left| {{\chi ^w}} \right|^2}$ as the worst-case passive beamforming gain achieved by using the beam flattening technique to solve problem \eqref{C5}. Then, we identify the minimum required number of APs, denoted by ${J_s}$, to achieve the fundamental squared power gain and further characterize ${\left| {{\chi ^w}} \right|^2}$ with respect to $N$ in the following theorem.
\begin{thm}
When $J \ge {J_s} \buildrel \Delta \over = \left\lceil {N\bar d\Delta _s^I} \right\rceil$, we have ${\left| {{\chi ^w}} \right|^2} \simeq 4{N^2}/{\pi ^2}$ as $N \to \infty$.
\end{thm}

\emph{Proof:} When $J \ge {J_s}$, its resulting angular span is $\left[ {{\Delta _{\min }},{\Delta _{\max }}} \right]$ and ${\Delta _{\max }} - {\Delta _{\min }} \le 1/\left( {N\bar d} \right)$. By setting ${\theta _n} =  - 2\pi \left( {n - 1} \right)\bar d\left( {{\Delta _{\min }} + 1/\left( {2N\bar d} \right)} \right)$. The worst case passive beamforming gain within $\left[ {{\Delta _{\min }},{\Delta _{\max }}} \right]$ can be derived as
\begin{align}\label{worst_pas_gain}
{\left| {{\chi ^w}} \right|^2} = \sum\limits_{n = 1}^N {{e^{ - i\pi \left( {n - 1} \right)/N}}}  = \frac{1}{{{{\sin }^2}\left( {\pi /\left( {2N} \right)} \right)}}\mathop  \simeq \limits^{\left( a \right)} \frac{4}{{{\pi ^2}}}{N^2},
\end{align}
where the approximation in ($a$) becomes tighter as $N \to \infty$ since $\mathop {\lim }\limits_{x \to 0} \left( {\sin x} \right)/x = 1$. Thus, the proof is completed.
$\hfill\blacksquare$

Theorem 1 indicates that the power scaling law of ${\cal O}\left( {{N^{2}}} \right)$ can be extended from the single-user case in \cite{wu2019beamforming} to the case for area coverage. Denote the worst-case received power of the \emph{D-MIMO empowered static IRS} architecture and the DIBF scheme as $\bar p_r^w\left( N \right)$ and $\bar p_r^d\left( N \right)$, respectively. Then, we have the following corollary to quantify the practical performance gap relative to the DIBF scheme in terms of the worst-case received power.
\begin{cor}
 When $J \ge {J_s}$, it follows that
 \begin{align}\label{conclusion}
\mathop {\lim }\limits_{N \to \infty } {\rm{ }}{{\bar p_r^w\left( N \right)} \mathord{\left/
 {\vphantom {{\bar p_r^w\left( N \right)} {\bar p_r^d\left( N \right)}}} \right.
 \kern-\nulldelimiterspace} {\bar p_r^d\left( N \right)}} \ge {4 \mathord{\left/
 {\vphantom {4 {{\pi ^2}}}} \right.
 \kern-\nulldelimiterspace} {{\pi ^2}}}.
\end{align}
\end{cor}

\emph{Proof:} Based on Theorem 1, we have
\begin{align}\label{worst_received_power}
\mathop {\lim }\limits_{N \to \infty } \bar p_r^w\left( N \right) &\mathop  \ge \limits^{\left( a \right)} \left( {{\gamma _1}M\mathop {\lim }\limits_{N \to \infty } {{\left| {{\chi ^w}} \right|}^2} + {\gamma _2}MN} \right)L \nonumber\\
&=\left( {{\gamma _1}\frac{4}{{{\pi ^2}}}M{N^2} + {\gamma _2}MN} \right)L,
\end{align}
where $L = \left( {\mathop {\min }\limits_{j \in {\cal J}} \rho _{1,j}^2} \right)\left( {\mathop {\min }\limits_{k \in {\cal K}} \mathop {\min }\limits_{{{\bf{u}}_k} \in {{\cal A}_k}} \rho _{2,k}^2\left( {{{\bf{u}}_{{k}}}} \right)} \right)$ accounts for the worst-case concentrated path-loss and ($a$) holds since $\mathop {\min }\limits_{x \in {\cal X}} f(x)g\left( x \right) \ge \mathop {\min }\limits_{x \in {\cal X}} f(x)\mathop {\min }\limits_{x \in {\cal X}} g\left( x \right)$. Then, $\bar p_r^d\left( N \right)$ can be expressed as $\bar p_r^d\left( N \right) = \left( {{\gamma _1}M{N^2} + {\gamma _2}MN} \right)L$. Thus, we obtain
\begin{align}\label{received_power_loss}
\frac{{\bar p_r^w\left( N \right)}}{{\bar p_r^d\left( N \right)}} \ge \frac{{{\gamma _1}4M{N^2}/{\pi ^2} + {\gamma _2}MN}}{{{\gamma _1}M{N^2} + {\gamma _2}MN}}\mathop  \ge \limits^{\left( b \right)} \frac{4}{{{\pi ^2}}},
\end{align}
where the equality in ($b$) holds when ${\gamma _1} = 1$ and ${\gamma _2} = 0$.
$\hfill\blacksquare$

Corollary 1 implies that the worst-case received power-loss of the \emph{D-MIMO empowered static IRS} architecture relative to the DIBF scheme is always less than 3.9 dB when the number of APs is sufficient.

\section{Numerical results}
We consider a three-dimensional Cartesian coordinate system, where the IRS is located at $\left( {0,0,10} \right)$ meter (m). The target area is assumed to be a rectangular centered at $\left( {150,0,0} \right)$ m. whose length and width are set as 100 m and 40 m, respectively. $J$ distributed APs are located on a circle centered at $\left( {0,0,0} \right)$ m with radius 10 m, whose specific locations satisfy ${\Phi _{r,j}} = {\Phi _{r,1}} + \left( {j - 1} \right)\Delta _s^I/J$ as indicated in Proposition 3. The target area is partitioned into $K = J$ disjoint subareas and subarea ${{\cal A}_j}$ satisfies ${\Phi _{t,j}}\left( {{{\bf{u}}_j}} \right) \in \left[ {\Phi _t^{\min } + \left( {j - 1} \right)\Delta _s^I/J,\Phi _t^{\min } + j\Delta _s^I/J} \right]$, ${{\bf{u}}_j} \in {{\cal A}_j}$. The pathloss exponents of both the AP-IRS and IRS-user channels are set to 2. The signal attenuation at a reference distance of 1 m is set to 40 dB. Unless otherwise stated, the remaining system parameters are set as: $M = 4$, ${P_{\max }} = 23$ dBm, and $\bar d = 0.5$.

\begin{figure}[!t]
 \centerline{\includegraphics[width=2.6in]{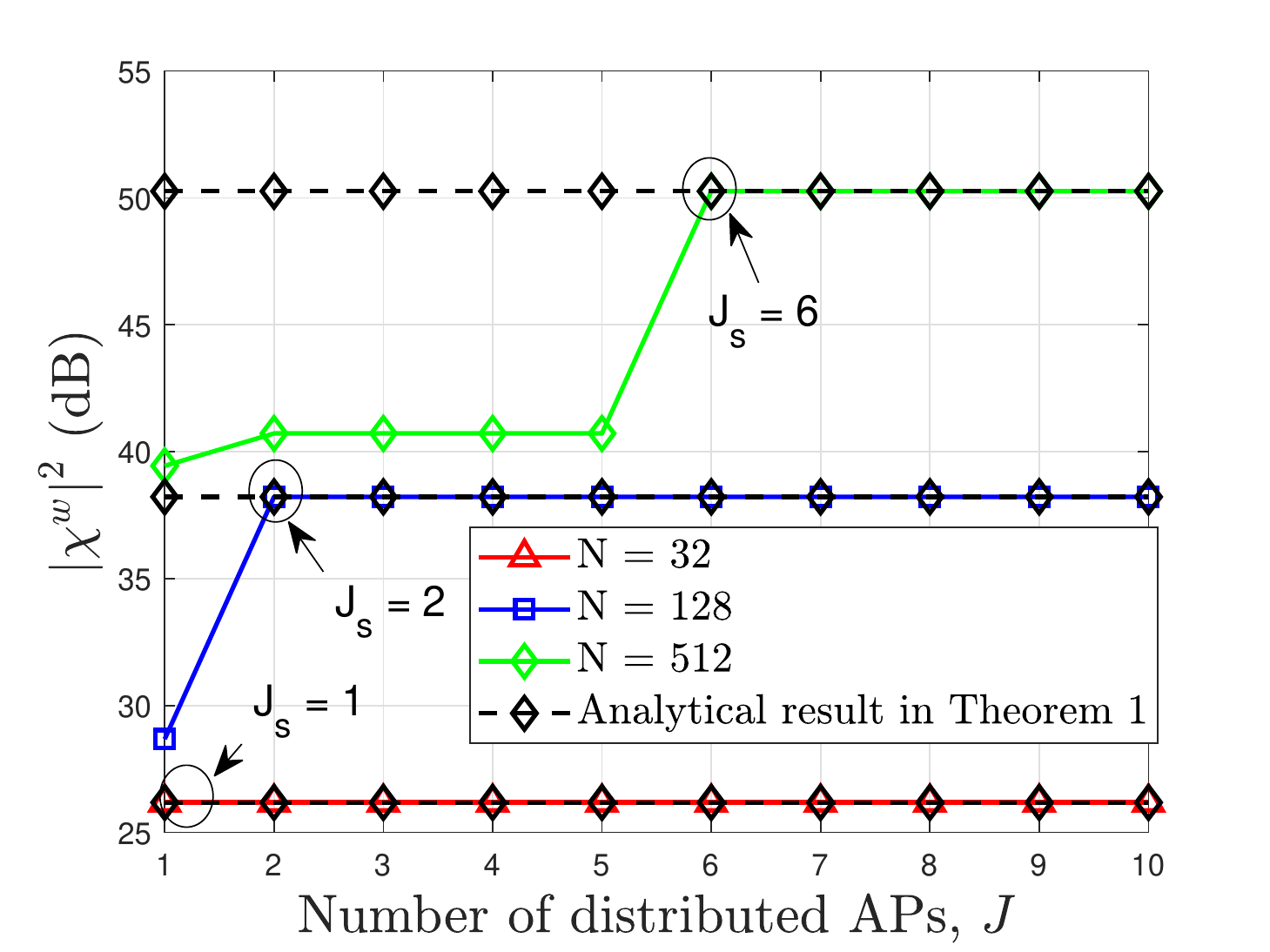}}
 \caption{Worst-case IRS passive beamforming gain versus $J$.}
 \label{passive_bf}
 \vspace{-8pt}
\end{figure}

In Fig. \ref{passive_bf}, we evaluate the worst-case passive beamforming gain versus $J$ under the different number of IRS elements. It is observed that the minimum required number of APs to achieve the fundamental squared power gain increases as $N$ becomes large, which is consistent with that the expression of ${J_s}$ shown in Theorem 1 is a increasing function with respect to $N$. This is because the beamwidth formed by the large IRS is narrow, which requires more APs to keep the resulting angular span falling into its beamwidth. Benefited by the diversity of different directions provided by distributed APs, the angular deviation is significantly reduced and thus substantial passive beamforming gain (more than 10 dB) over the case of $J = 1$ can be achieved. Additionally, the value of the achieved passive beamforming gain is near to $4{N^2}/{\pi ^2}$, which agrees with our analysis in Theorem 1.

\begin{figure}[!t]
 \centerline{\includegraphics[width=2.6in]{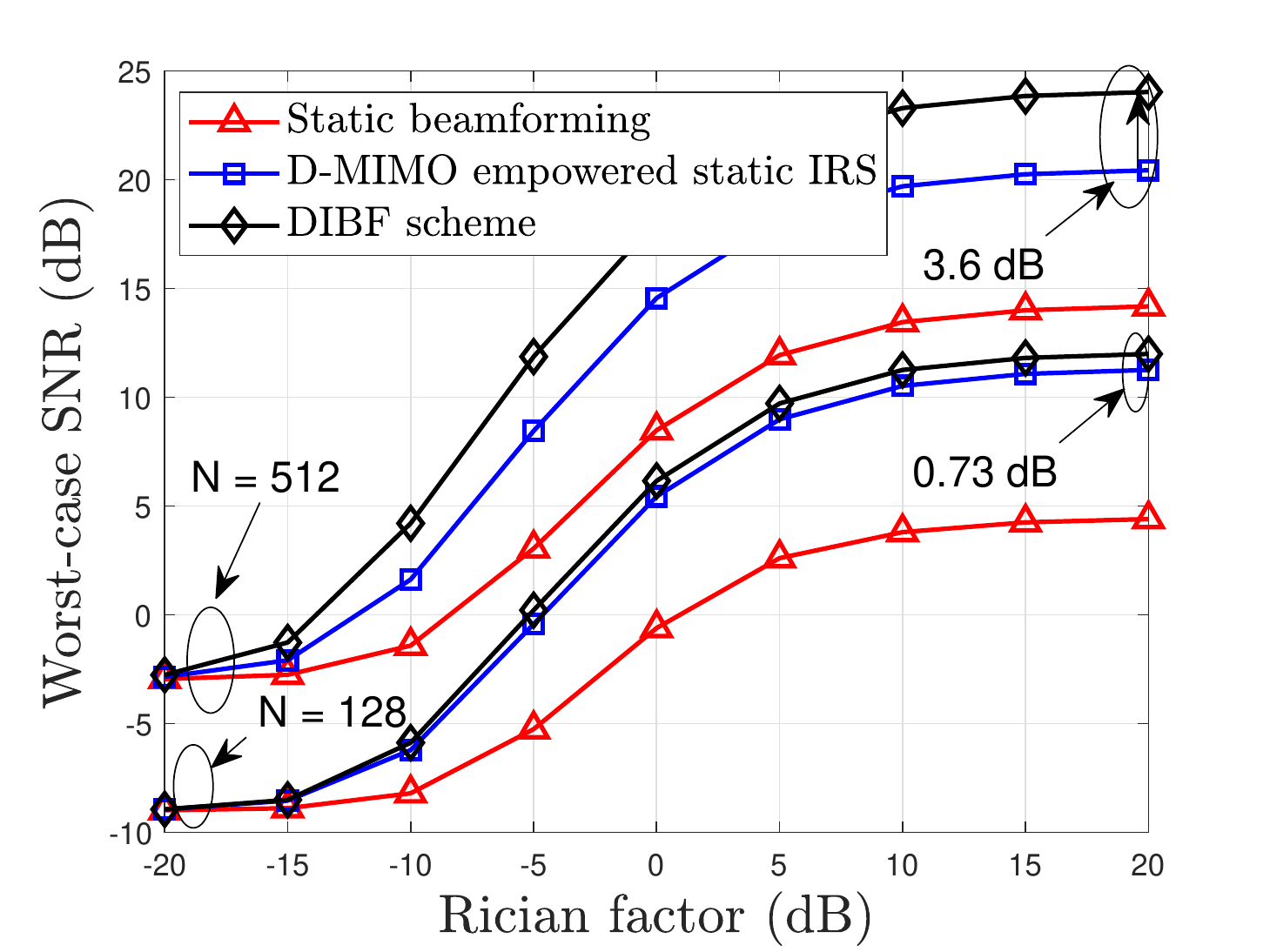}}
 \caption{Worst-case SNR versus Rician factor with noise power $ - 90$ dBm .}
 \label{worst_case_snr}
 \vspace{-8pt}
\end{figure}

In Fig. \ref{worst_case_snr}, we show the worst-case SNR versus the Rician factor by normalizing the resulting received power by the noise power $-90$ dBm. The results of the proposed \emph{D-MIMO empowered empowered IRS} architecture are obtained under the condition of $J = J_s$. The DIBF scheme and the static IRS beamforming scheme (with $J = 1$) \cite{9351782} are also considered for performance comparison. It is observed that the performance gain of our proposed scheme over the static IRS beamforming scheme becomes more pronounced as Rician factor increases, which implies the significance of deploying the IRS to create LoS links with both the target-area and the AP. Moreover, the performance-loss of the proposed architecture compared to the DIBF scheme is observed to be 0.73 dB and 3.6 dB for $N = 128$ and $N = 512$, respectively. The resulting performance-loss is lower than 3.9 dB, which validates the analysis in Corollary 1.

\section{Conclusion}
For the static IRS enhanced wireless coverage, we proposed a novel \emph{D-MIMO empowered static IRS} architecture to achieve dynamic beamforming gain with only a single set of static IRS beamforming pattern. By exploiting the spatial direction of multiple APs, the target area is partitioned into multiple subareas and each subarea is covered by an assigned AP. We aim to maximize the worst-case average received power over all locations in the target area by jointly optimizing a single set of IRS beamforming pattern and AP-subarea association. An efficient algorithm was proposed to solve it by decoupling the AP-subarea association optimization and IRS beamforming design. Moreover,
we demonstrated both analytically and numerically the achieved performance gains of the proposed design over the conventional static beamforming design.

%
%

\bibliographystyle{IEEEtran}

\bibliography{IEEEabrv,myref}
\end{document}